\documentclass{article}

\usepackage{listings}
\usepackage{arxiv}
\usepackage{adjustbox}
\usepackage[utf8]{inputenc} % allow utf-8 input
\usepackage[T1]{fontenc}    % use 8-bit T1 fonts
\usepackage{hyperref}       % hyperlinks
\usepackage{url}            % simple URL typesetting
\usepackage{booktabs}       % professional-quality tables
\usepackage{amsfonts}       % blackboard math symbols
\usepackage{nicefrac}       % compact symbols for 1/2, etc.
\usepackage{microtype}      % microtypography
\usepackage{lipsum}		% Can be removed after putting your text content
\usepackage{graphicx}
\usepackage{natbib}
\usepackage{doi}
\usepackage{bm}
\usepackage{rotating}
\usepackage{tikz}
\usepackage{caption,color,float}

\usepackage[bb=px]{mathalfa} 
\usepackage[varvw]{newtxmath}       % selects Times Roman as basic font
\usepackage{amsmath}
\mathchardef\mhyphen="2D % Define a "math hyphen"
\usepackage{booktabs}
\usepackage{multirow}
\usepackage{comment}
\usepackage{diagbox}
\usepackage{soul}
\usepackage{fancyhdr}
\usepackage{wrapfig}
\usepackage{graphicx}

\usepackage{subcaption}

\usepackage{subcaption}

\title{AI-driven Uncertainty Quantification \& Multi-Physics Approach to Evaluate Cladding Materials in a Microreactor}

\author{
    {Alex ~Foutch} \\
	Nuclear Engineering and Radiation Science\\
	Missouri University of Science and Technology\\
	Rolla, MO 65409, USA \\
    \And
    {Kazuma ~Kobayashi} \\
    Nuclear, Plasma \& Radiological Engineering \\
    University of Illinois at Urbana-Champaign \\
    Urbana, IL 61801, USA \\
    \texttt{kazumak2@illinois.edu} \\
    \And
    {Ayodeji ~Alajo} \\
	Nuclear Engineering and Radiation Science\\
	Missouri University of Science and Technology\\
   Rolla, MO 65409, USA \\
    \And
    {Dinesh ~Kumar} \\
    Department of Mechanical Engineering\\
    University of Bristol, Bristol\\
    BS8 1TR, UK
    \And
    {Syed Bahauddin ~Alam} \\
    Nuclear, Plasma \& Radiological Engineering \\
     National Center for Supercomputing Applications \\
    University of Illinois at Urbana-Champaign \\
    Urbana, IL 61801, USA \\
    \texttt{alams@illinois.edu} \\
}

\renewcommand{\headeright}{Preprint}
\renewcommand{\undertitle}{Preprint}
\renewcommand{\shorttitle}{\textit{Virtual Sensing-Enabled Digital Twin Framework for Nuclear Systems
}}

\hypersetup{
pdftitle={A template for the arxiv style},
pdfsubject={q-bio.NC, q-bio.QM},
pdfauthor={David S.~Hippocampus, Elias D.~Striatum},
pdfkeywords={First keyword, Second keyword, More},
}

\begin{document}
\maketitle

\begin{abstract}
{ \small

The pursuit of enhanced nuclear safety has spurred the development of accident-tolerant cladding (ATC) materials for light water reactors (LWRs). This study investigates the potential of repurposing these ATCs in advanced reactor designs, aiming to expedite material development and reduce costs. The research employs a multi-physics approach, encompassing neutronics, heat transfer, thermodynamics, and structural mechanics, to evaluate four candidate materials (Haynes 230, Zircaloy-4, FeCrAl, and SiC-SiC) within the context of a high-temperature, sodium-cooled microreactor, exemplified by the Kilopower design.  While neutronic simulations revealed negligible power profile variations among the materials, finite element analyses highlighted the superior thermal stability of SiC-SiC and the favorable stress resistance of Haynes 230. The high-temperature environment significantly impacted material performance, particularly for Zircaloy-4 and FeCrAl, while SiC-SiC's inherent properties limited its ability to withstand stress loads. Additionally, AI-driven uncertainty quantification and sensitivity analysis were conducted to assess the influence of material property variations on maximum hoop stress. The findings underscore the need for further research into high-temperature material properties to facilitate broader applicability of existing materials to advanced reactors. Haynes 230 is identified as the most promising candidate based on the evaluated criteria.
}

\end{abstract}

%------------------------------------------------------------------------------
%
%------------------------------------------------------------------------------
{ 
\section{Introduction}
\label{sec:intro}

The 2011 Fukushima Daiichi disaster sparked a renewed focus on nuclear innovations for accident prevention, particularly targeting the materials used for uranium fuel and fuel rod claddings \cite{almutairi2022weight}. These accident-tolerant fuels and claddings (ATFs and ATCs) are primarily being developed for light water reactors (LWRs) \cite{terrani.2018}. ATCs are engineered to resist oxidation, a critical feature in LWR accident scenarios where increased steam in the coolant system could otherwise accelerate heat generation and hasten core degradation \cite{ART-ONE}. Yet, their benefits extend beyond oxidation resistance, offering additional advantages for accident tolerance that this paper explores.

Zirconium-based claddings have been the main choice for uranium fuel pellets in LWRs since the 1950s \cite{terrani.2018,alam2019assembly} because of their low thermal neutron absorption cross section and resistance to corrosion. Their adoption in the nuclear industry expanded in the 1960s, establishing zirconium alloys as the preferred cladding material for LWRs globally. However, at high temperatures in water, zirconium alloys exhibit undesirable traits like increased corrosion and material degradation, limiting their in-core lifespan. Furthermore, in severe accidents, the self-accelerating exothermic steam-zirconium reaction at above approximately 1200°C is a major drawback, generating substantial heat and hydrogen gas \cite{hofmann1999current,steinbruck2011air}. The flammability of hydrogen can lead to explosions upon contact with hot surfaces (520-720°C) \cite{9hollnagel2013fukushima}. Hydrogen-initiated explosions due to steam-cladding reactions contributed to the Fukushima accident's failure to build containment. This event prompted the industry to re-evaluate fuel and cladding designs for improved accident tolerance. As one of the most common Zr alloys, Zircaloy-4 was chosen for this paper. Although it is not designated an ATC, Zircaloy-4 will provide a baseline for the other candidates. 

Iron-based nuclear claddings have been used since 1951, where austenitic stainless steel was deployed in the Experimental Breeder Reactor I \cite{terrani.2018}. Survey tests of Fe-based alloys found the high-temperature oxidation resistance of FeCr, FeAl, and FeCrAl made them desirable. Of these, FeCrAl was chosen to move forward and has undergone significant testing. Normal operational and anticipated operational occurrence behavior are expected to be superior to Zr-based cladding, but based mainly on its enhanced oxidation resistance \cite{terrani.2018}. FeCrAl will be one of the candidate claddings.

Silicon Carbide (SiC)’s use in the nuclear industry reaches far back, but high-strength fiber-reinforced composites (SiC-SiC, as opposed to bulk SiC) date to research in the 1970s. The technology for its production has improved since, and owing to its exceptional oxidation resistance, it is likewise deemed an ideal accident-tolerant cladding \cite{terrani.2018}. Its status as a ceramic also lends itself to different material behaviors from metals, which may prove noteworthy for this analysis. SiC-SiC will thus be another candidate cladding. Also, Haynes 230, the original choice for the heat pipes and ring clamps, will be included in the simulations as another baseline for Kilopower Reactor Using Stirling Technology (KRUSTY) benchmarking.

\begin{table}[tb]
  \caption{Candidate Cladding Material Properties}
  \label{tbl:Cladding-Props}
  \begin{center}
  \resizebox{\columnwidth}{!}{\begin{tabular}{l r r r r}
  \hline
   & Haynes 230 \cite{Haynes-Brochure} & Zircaloy-4 \cite{ATI-Zirc} & SiC-SiC \cite{Handbook-SiC,NEI-SiC} & FeCrAl \cite{Therm-FeC,Handbook-FeC} \\
  \hline
  Young's Modulus ($GPa$) & 159 & 99.3 & 240 & 135 \\
  Poisson's Ratio & 0.34 & 0.37 & 0.191 & 0.3 \\
  Thermal Expansion Coefficient ($1/K$) & 15.2e-6 & 6e-6 & 5.5e-6 & 13.4e-6 \\
  Thermal Conductivity ($W/m*K$) & 24.4 & 21.5 & 68.85 & 22 \\
  Specific Heat Capacity ($J/kg*K$) & 595 & 285 & 1200 & 710 \\
  Density ($kg/m^{3}$) & 8970 & 6550 & 2700 & 7150 \\
  Melting / Subl. Point ($K$) & 1574 & 2123 & 2700 & 1773 \\
  \hline
  \end{tabular}}
  \end{center}
\end{table}

The oxidation resistance of these candidates will not be relevant to this analysis. Rather, it is their thermodynamic and mechanical behaviors that will be examined. The properties used for the four candidates are compiled in Table \ref{tbl:Cladding-Props}. Also included is the melting point/sublimation point (for SiC-SiC), a common failure limit against which the candidates' performance can be measured. Although not a candidate cladding, the maximum temperature of the U-8Mo (8 $wt\%$ $Mo/(U+Mo)$) fuel will be compared against its melting point, 1408 $\rm K$, as another performance metric \cite{creasy.2011}.

\begin{equation}\label{eq:Fraction}
    Variable \ Fraction = \frac{Maximum \ Measure \ of \ Variable}{Limit \ of \ Variable}
\end{equation}

\begin{figure}[tb]
  \begin{center}
  \includegraphics[width=4.25in]{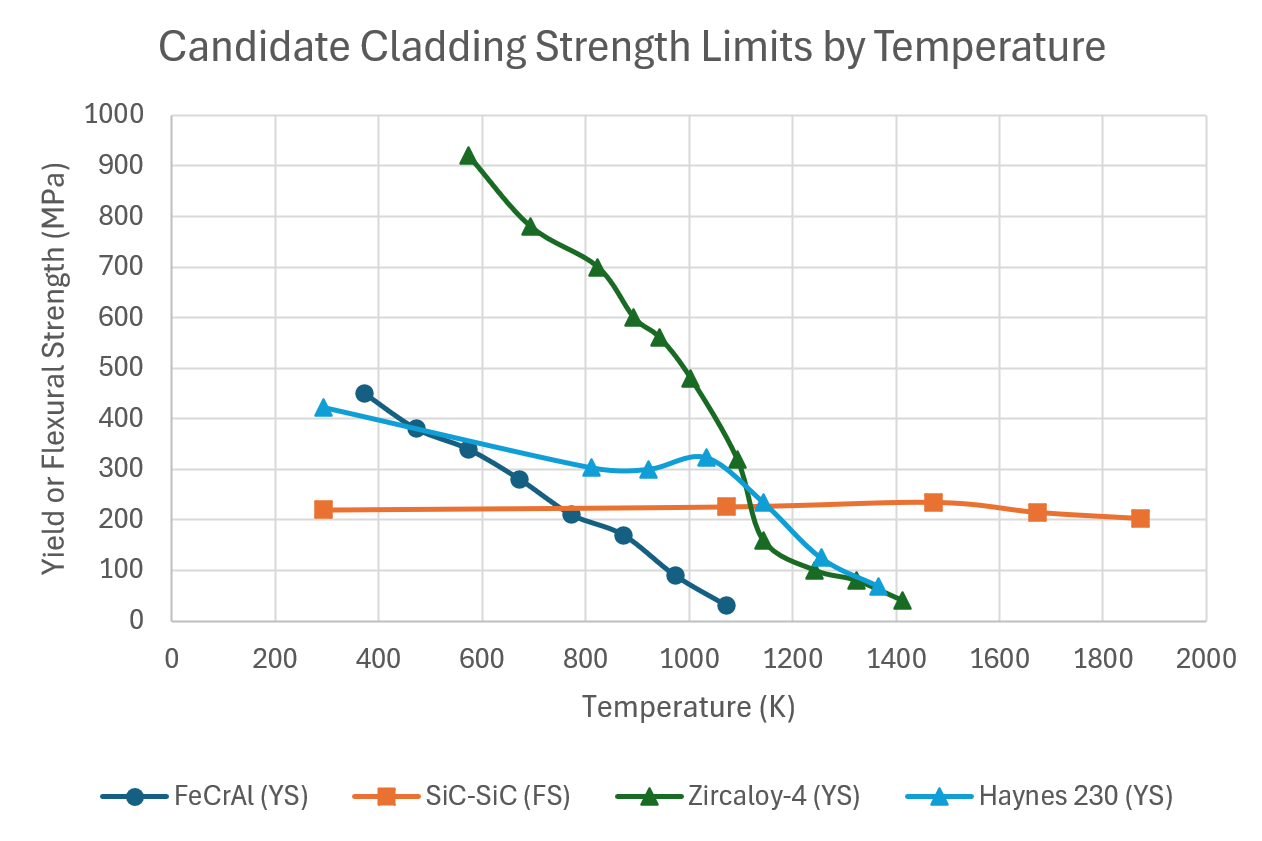}
  \end{center}
  \caption{Yield strength (YS) or flexural strength (FS) for each candidate cladding \cite{Handbook-FeC,xu.2019,PNNL-Zirc,Haynes-Brochure}.}
\label{fig:Cladding-Limits}
\end{figure}%

Another failure limit, yield strength (YS) / flexural strength (FS) (for SiC-SiC), is plotted against temperature in Figure \ref{fig:Cladding-Limits}. If temperatures exceed the limits of this data, the closest data point will simply be used (so 1800 K FeCrAl would use 30 MPa, the same as 1100 K FeCrAl). This failure analysis will be further elaborated on in Section \ref{sec:MOOSE}. The measurement for any failure limit will be their `fraction,' Equation \ref{eq:Fraction}, each named after the environment, component, and operating variable. For instance, a comparison of maximum fuel temperature and fuel melting point during NOC (Normal Operating Conditions) would be the `NOC fuel temperature fraction. Such new materials take long periods to be fabricated, tested and implemented. The landscape of nuclear power production is also changing with the advent of advanced and small modular reactor designs that are very different from existing LWRs.

To save future development time on claddings for advanced / small modular reactors, this paper seeks to examine the performance of candidate ATCs in such an environment. Compared to LWRs, this will impact the temperatures, pressures, irradiation, and stresses experienced by cladding. This includes experiences at NOC and during a loss of coolant accident (LOCA). This will mainly measure the materials' proximities to failure limits during simulation, comparing the candidates' fractions of performance vs limit between each other and determining which is least likely to reach those limits (or which already exceeds them).

A LOCA occurs when coolant is blocked or diverted and cannot flow through the reactor core. This leads to higher temperatures and pressures than the cladding is designed for and may even lead to dangerous levels of thermal expansion and/or eventual rupture, releasing irradiated coolant and potentially a meltdown if coolant flow cannot be restored. Of the common cladding failure modes, LOCAs were chosen for analysis due to their relation to the aforementioned design basis of ATCs \cite{khattak.2019}. Since ATCs are specifically engineered to withstand extreme conditions, including high-temperature oxidation and hydrogen production in accident scenarios, their performance under LOCA conditions serves as a key validation metric for their effectiveness. They also present a very real opportunity for extreme conditions in which to place the candidate claddings. And it is the very same incident that cascaded into the disaster at Fukushima \cite{Fukushima-WN}.

The research gap this article addresses involves evaluating and applying accident-tolerant cladding (ATC) materials originally developed for light water reactors (LWRs) in advanced reactor environments such as molten salt-cooled microreactors. While these materials have been extensively tested and developed for LWRs, their performance under the different conditions present in advanced reactors, such as higher temperatures and different types of coolants (e.g., sodium), has not been thoroughly investigated. The article seeks to fill this gap by using a multiphysics analysis combined with machine learning-based uncertainty quantification to assess the feasibility of using these ATC materials (e.g., Haynes 230, Zircaloy-4, FeCrAl, and SiC-SiC) in such advanced reactor designs, specifically using the Kilopower microreactor as a case study. The study aims to provide insights into these materials' mechanical and thermodynamic behavior under NOC and LOCA, thus offering recommendations for their potential application in advanced nuclear reactors. The research intends to save time and development costs by repurposing existing materials for new reactor designs rather than developing entirely new materials from scratch.\\

The primary objectives of the paper are:

\begin{enumerate}
    \item To explore whether ATCs, designed for LWRs, suit advanced reactor environments like Kilopower, aiming to cut development time and costs.
    
    \item To explore the performance of various candidate ATCs (Haynes 230, Zircaloy-4, FeCrAl, and SiC-SiC) under NOC and LOC scenarios in a high-temperature, sodium-cooled microreactor environment. The study evaluates the materials' resistance to temperature changes, stress levels, and proximity to failure limits.
    
    \item To apply a multi-physics approach—integrating neutronics, heat transfer, thermodynamics, and continuum mechanics—for a holistic view of cladding behavior.
    
    \item To assess the impact of uncertainties in material properties on the simulation results using uncertainty quantification (UQ) and sensitivity analysis (SA). The goal is to identify which material properties have the most significant influence on the cladding's performance and would require further experimental investigation.
\end{enumerate}

\section{Reactor System}

\subsection{Kilopower / KRUSTY}
\label{subsec:KP}

In terms of existing concepts, the Kilopower reactor developed at NASA Glenn is a good base. It is a fission surface power (FSP) microreactor designed to be part of a lunar colony’s electric grid \cite{poston.2020}. This necessitates many differences from commercial LWRs, detailed in Table~\ref{tbl:LWR-vs-KP}. 

%Finally the tables, Table~\ref{tbl:tbl01} illustrates the syntax of a basic table. %
\begin{table}[tb]
  \caption{Characteristics of a Typical LWR vs Kilopower}
  \label{tbl:LWR-vs-KP}
  \begin{center}
  \begin{tabular}{l l l}
  \hline
   & Typical LWR \cite{NPRC,LWRs} & Kilopower \cite{poston.2020} \\
  \hline
  Fuel & UO2 & U-8Mo \\
  Enrichment ($\%$) & 4 &  93  \\
  Operating Power (MWe) & 700 & 1-10 \\
  Coolant & Light water (two-phase) & Sodium (two-phase) \\
  Cladding & Zirconium alloy & Haynes 230 \\
  \hline
  \end{tabular}
  \end{center}
\end{table}

More specifically, Kilopower's fuel is monolithic, made of three separately-wrought cylindrical blocks of U-8Mo fuel, coming together at 25 $cm$ long axially. A B4C (Boron Carbide) control rod is inserted into a central hole in the fuel. It is the reactor's only moving part. Eight heat pipes in angular symmetry are inserted within outer gaps in the fuel. The pipes themselves are made of Haynes 230, with a thin inner layer of nickel as a wick against in-flowing two-phase sodium coolant and a thin outer layer of copper foil to prevent mass diffusion/transfer between the U-8Mo and the Haynes 230 at operational temperatures. Seven equally spaced ring clamps surround the fuel and heat pipes, providing both thermal and structural coupling. All these components are held together via an interference fit - heating the clamps to high temperature, expanding them, and allowing them to cool around the core until they are all firmly pressed together \cite{poston.2020}.

This compact design was chosen to minimize volume/mass during transport into and throughout space and to provide modularity, allowing it to be combined with other Kilopowers to satisfy a power system's needs. Hence, its use of highly enriched uranium (HEU) allowed for higher power density from the fuel. Heat pipes also eliminated the need for pumps to drive the coolant flow \cite{poston.2020}, further condensing the load. These design decisions set very different conditions for its structural components to satisfy. For instance, an LWR often has its uranium fuel encased in cladding, a protective outer layer, separated with a small gas gap to allow thermal expansion of the uranium fuel. This design has no such precaution, meaning that the Kilopower fuel will not only experience a greater rate of heat transfer from it to its heat pipes, but the rest of the core must be able to sustain the fuel's thermal expansion directly.

Kilopower is a particularly helpful benchmark for this paper, having been recently and extensively tested \cite{poston.2020,gibson.2018}. Results of the KRUSTY Nuclear Ground Test conducted on March 21, 2018, are readily available. KRUSTY slightly differs from the default Kilopower `flight' setup. These changes are negligible for this analysis, but the KRUSTY configuration will be prioritized unless specified otherwise. The test configuration adds several outer layers not in the initial Kilopower design, pictured in Figure \ref{fig:KRUSTY-config}. Multi-layer insulation (MLI), composed of molybdenum foil, prevents significant heat loss from the core. At the same time, a stainless steel vacuum enclosure simulates the environment of outer space and the lunar surface. Both designs include a large, cylindrical BeO (Beryllium Oxide) reflector/shield around the core apparatus. Although KRUSTY's is broken into a platen, a shim, and an axial section, they are roughly of equal shape and size. A thin stainless steel sleeve inside the shield ensures alignment and prevents contact with the vacuum can \cite{poston.2020}.

\begin{figure}[tb]
  \begin{center}
   \includegraphics[width=5.75in]{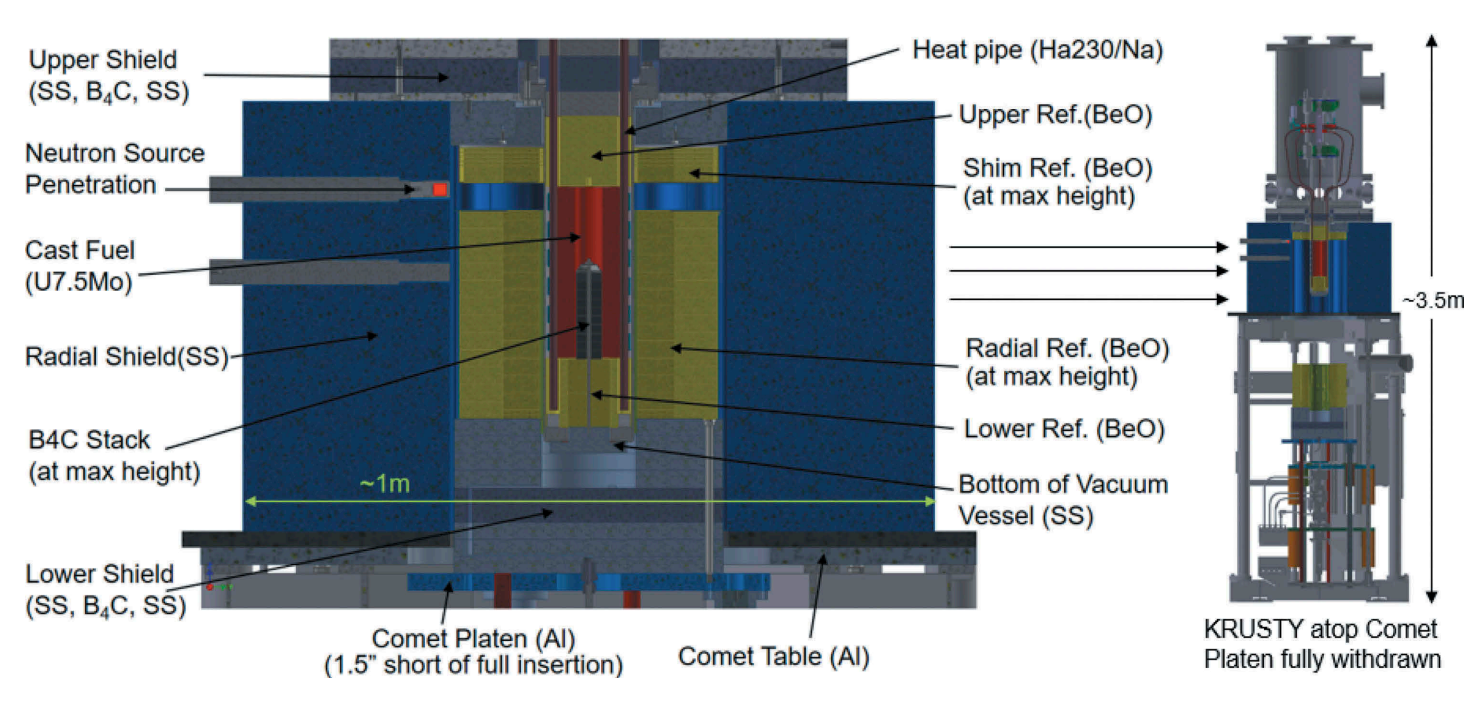}
  \end{center}
  \caption{KRUSTY configuration \cite{poston.2020}.}
\label{fig:KRUSTY-config}
\end{figure}%

If it were simply a concept tested only through modeling and simulation (M\&S), unforeseen factors, such as unknown chemical interactions under high temperatures, pressures, and irradiation, might force redesigns. Having verification of this design's viability lends Kilopower an authenticity for what can be expected of some advanced, non-LWR reactors, such as its choice of heat pipes, sodium coolant, a monolithic core, etc.

\subsection{Kilopower Applications}
\label{subsec:KPapp}

On applying ATCs to Kilopower, this reactor is not designated as having a ‘cladding’ \cite{poston.2020}. However, the general purposes of nuclear fuel cladding are to:

\begin{itemize}
  \item Transfer heat and prevent corrosion between coolant and fuel
  \item Contain swelling of fuel pellets
\end{itemize}

Looking at Kilopower’s design, these roles are achieved by its eight heat pipes and a series of ring clamps that surround the core – all of which are made of Haynes 230, a nickel-chromium-tungsten-molybdenum alloy designed for use in aerospace and energy industries \cite{Haynes-Brochure,poston.2020}. In its simulations, this paper will directly replace Haynes 230 with the candidate claddings to perform a parametric analysis. It is understood that Haynes 230 may have been chosen for heat pipes and ring clamps for reasons outside the scope of this paper, such as cost, ease of manufacture, or performance in other areas. As such, this analysis will not conclude whether the candidates are a good fit for Kilopower specifically; rather, it will compare their performance in conditions typical of this reactor environment (HEU, micro, etc.), using Kilopower as a benchmark. The reported results of KRUSTY will be used to define operational parameters. This means the fuel will generate 5 kW of heat and maintain an average temperature of 800$^{\circ}$C. Two-phase Sodium coolant will be simulated through the heat pipes, with inlet and outlet temperatures based on the Stirling converter cold and hot ends of 70$^{\circ}$C and 660$^{\circ}$C. The core will be in a vacuum, instead of air, and surrounded by a BeO shield %\cite{gibson.2018}. 

This work is designed to be as accurate to the conditions expected in this reactor environment as is reasonably necessary. Certain factors have been approximated or cut for expediency because the impacts on the results are deemed negligible or outside this paper’s scope. These will be noted throughout the paper.

\section{Methodology \& Results}

\subsection{Power Profiling}
\label{subsec:MCNP}
\label{subsec:MPA}

\begin{figure}[tb]
  \begin{center}
  \includegraphics[width=3.25in]{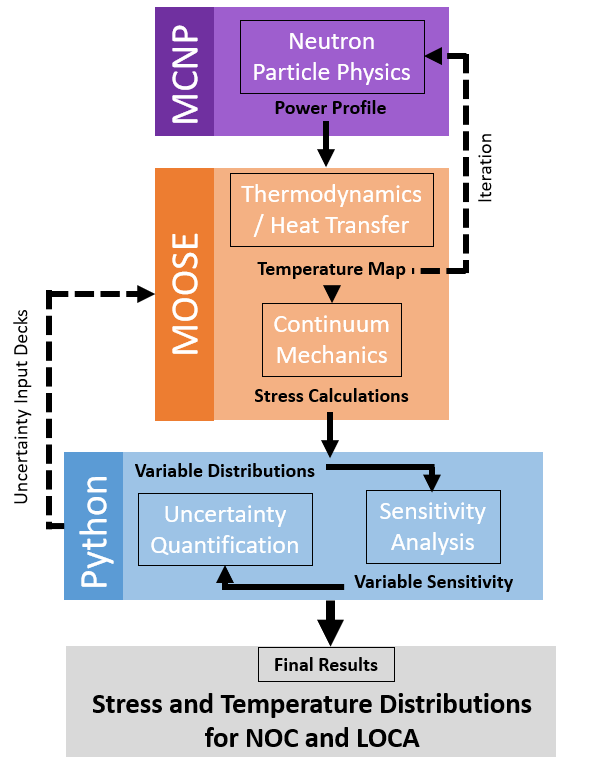}
  \end{center}
  \caption{The combined M\&S approach to model Kilopower performance and failure modes.}
\label{fig:Multi-Phys-Plan}
\end{figure}%

The analysis done on Kilopower will be multi-physics, meaning multiple fields of modeling and simulation (M\&S) will be used and coupled together, with the goal of providing more holistic results. This is illustrated visually in Figure \ref{fig:Multi-Phys-Plan}. In this case, neutron transport, heat transfer, thermodynamics, and continuum mechanics. This approach provides a throughline to arrive at the desired results: performance and failure modes of the candidate claddings, based on the measure of temperature and stress levels they experience.

Monte Carlo N-Particle Transport (MCNP) is a probabilistic software developed by Los Alamos National Laboratory (LANL) and designed to track simulated particles through user-defined geometry \cite{MCNP-Manual}. It can simulate complex geometries with sufficient customization to replicate the Kilopower reactor. It was run remotely on The Foundry, Missouri University of Science \& Technology (S\&T)’s High Performance Computing (HPC) cluster.

\label{subsec:MCNPM}

The M\&S effort began here, as the initial assumption was that while a thermodynamic/heat transfer model could provide valuable insights, its direct impact on the neutronics \cite{alam2019neutronic}, particularly on power peaking calculations, would be limited. Specifically, the axial power profile of the Kilopower fuel was measured using fission rate measurements from MCNP. While the core has a total thermal power rating, knowing which areas have the highest heat density created a more accurate heatmap of the core for thermal analysis. Calculating the axial power profile (APP) from the fission rate involves converting it to a unitless representation of the power shape. To find this shape, each axial fission rate was divided by the average fission rate across the axial length of the core.

\begin{figure}[tb]
  \begin{center}
   \includegraphics[width=2.75in]{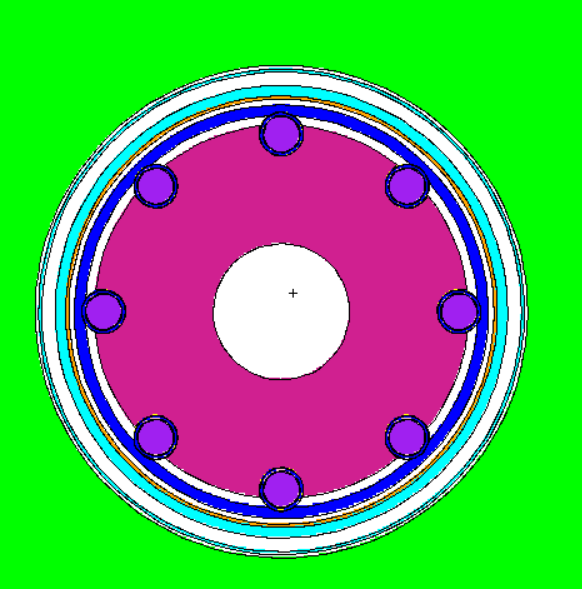}
   \includegraphics[width=3.15in]{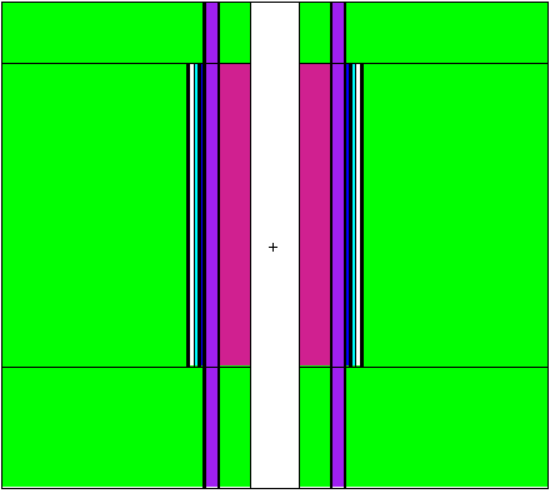}
  \end{center}
  \caption[KRUSTY core recreated in MCNP.]{KRUSTY core recreated in MCNP, viewed at the radial-angular and radial-axial planes. From inner to outer, the central layers are fuel (fuchsia), singular ring clamp (blue), MLI (orange), vacuum can and steel sleeve (teal), and BeO shield (green). White represents a void/vacuum. The eight heat pipes (blue) are filled with a customizable Sodium material (purple) and have inner nickel wick and outer copper foil layers (unseen).}
\label{fig:MCNP-XY-YZ}
\end{figure}%

Taking baseline temperatures and material properties (density and isotopic composition), the Kilopower core and surrounding KRUSTY configuration were constructed. The reflector was deemed a boundary where no particles that escaped it would be relevant, so no geometry outside of it was rendered, and particles beyond this boundary were `killed.' The rendered geometry can be seen in Figure \ref{fig:MCNP-XY-YZ}. The six ring clamps were combined into one shape to prevent local peaking at each of the five gaps from moderated neutrons entering the core directly from the BeO reflector, and thus make the power profile easier to perform regression upon. In its original testing, these represented only a 1\% deviation and were deemed by the Kilopower team as negligible \cite{poston.2020}.

%\section{Results}
\label{subsec:MCNPR}

\begin{figure}[tb]
  \begin{center}
   \includegraphics[width=3.75in]{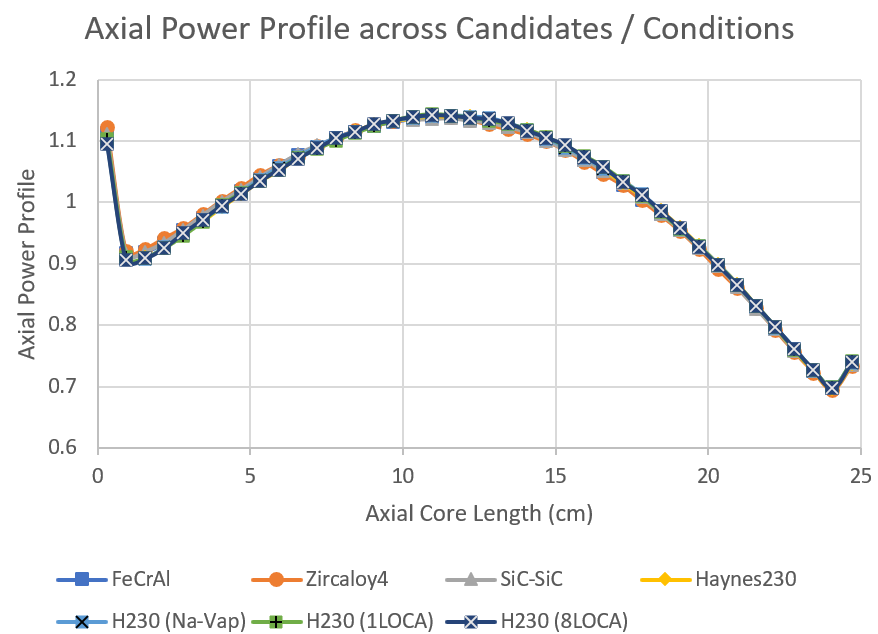}
   \includegraphics[width=3.75in]{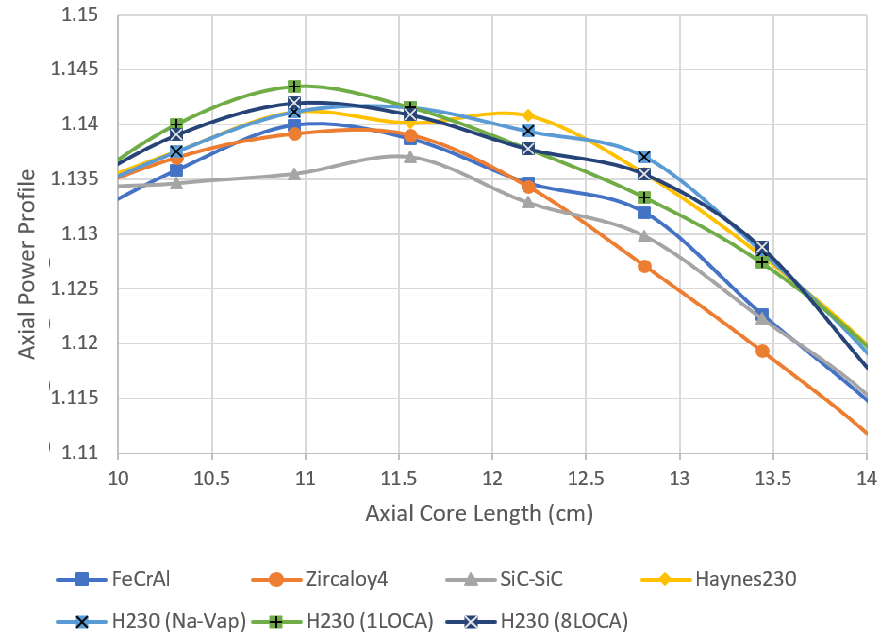}
  \end{center}
  \caption[Axial power peaking of the Kilopower fuel block.]{Axial power peaking of the Kilopower fuel block - both the entire core and magnified for clarity.}
\label{fig:FR-APP}
\end{figure}%

Various runs tested the impact of candidate cladding materials, sodium coolant composition, and LOCA occurrences on the APP results. These cases include: 1LOCA (a scenario where a single heat pipe experiences a loss of coolant), 8LOCA (a more severe accident where all eight heat pipes fail, leading to a complete loss of coolant) and Na-Vap (a case where sodium exists primarily in its vapor phase, affecting heat transfer properties). In LOCA scenarios, the power profile is assumed to remain steady, reflecting the reactor’s fission rate distribution before any transient effects occur. However, slight deviations from the simple Haynes 230 calculation arise due to differences in material properties and thermal feedback effects, which subtly alter the neutron flux distribution. All are collected in Figure \ref{fig:FR-APP} and display minimal deviation among power shapes. Because the peaks at the edges represented a thin margin of either power profile, these data points were deemed insignificant enough to be removed before performing linear regression. To reiterate, this paper seeks to recreate comparable conditions in this type of microreactor, not recreate Kilopower or KRUSTY one-for-one. Even the Kilopower team deemed such peaks to be insignificant to heat transfer calculations \cite{poston.2020}.

\begin{figure}[tb]
  \begin{center}
   \includegraphics[width=3.75in]{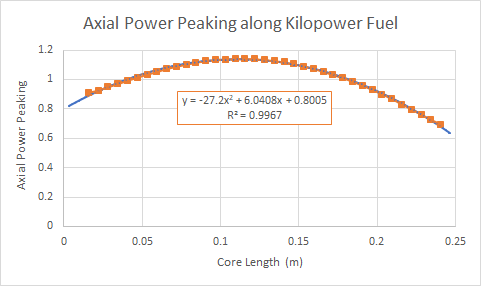}
  \end{center}
  \caption{APP of the entire Kilopower fuel block (orange) and the extrapolated equation to represent it (blue).}
\label{fig:FR-APP-Final}
\end{figure}%

Equation \ref{eq:APP} is the APP of the Kilopower fuel for Haynes 230, drawn from Figure \ref{fig:FR-APP-Final}. Its range is from 0 to 0.25 $m$ and will be multiplied by Power Density in the MOOSE model.\begin{equation}\label{eq:APP}
    APP = -27.2x^2 + 6.0408x + 0.8005
\end{equation}

\subsection{Thermal and Stress Analysis}
\label{sec:MOOSE}

Multiphysics Object Oriented Simulation Environment (MOOSE) is a finite element framework developed by the Idaho National Laboratory (INL), used by this paper for its heat transfer and continuum mechanics capabilities \cite{MOOSE-Main}.  

%\subsection{Methodology}
\label{subsec:MOOSEM}

Due to angular symmetry within the core, only a 1/4 slice was modeled. Additionally, the nature of this analysis dictated only the ring clamp experiencing the highest temperatures needed to be modeled, as it would be the first to experience stress failure from thermal expansion. The ring clamps positioned along the axial length of the core were selected for modeling (as shown in Figure \ref{fig:KP-1-4}) because it corresponds to the highest axial power peaking region, making it the most critical for assessing thermal and mechanical stresses.

\begin{figure}[tb]
  \begin{center}
   \includegraphics[width=3.75in]{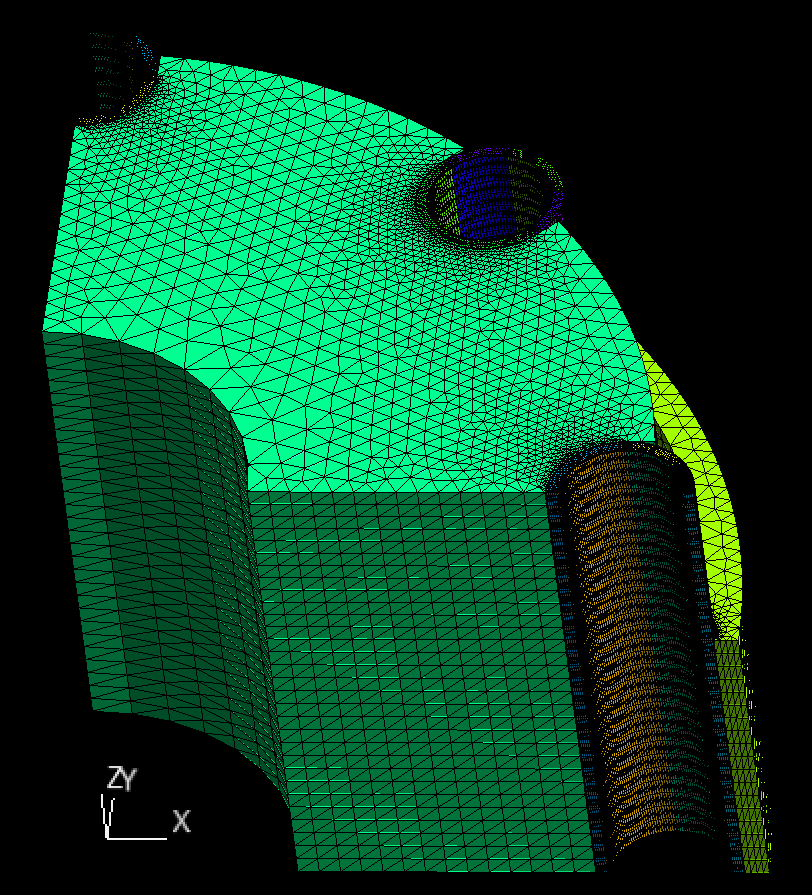}
  \end{center}
  \caption{1/4 section of the Kilopower core and RCD (fuel, heat pipe, ring clamps).}
\label{fig:KP-1-4}
\end{figure}%

\begin{table}[tb]
  \caption{Maximum Hoop and Temperature for Candidate Cladding (NOC)}
  \label{tbl:T_Hoop-Props1}
  \begin{center}
  \resizebox{\columnwidth}{!}{\begin{tabular}{l r r r r}
  \hline
   & Haynes 230  & Zircaloy-4  & SiC-SiC  & FeCrAl  \\
  \hline
  Fuel Max Temp (K) & 1111.15 & 1111.94 & 1104.98 & 1111.81   \\
  Clad Max Hoop (Pa) & 236210971.00 & 671990320.90 & 1023613044.00 & 298991081.90  \\
  Clad Max Temp (K)  & 1083.33 & 1084.13 & 1076.75 & 1084.00  \\
  \hline
  \end{tabular}}
  \end{center}
\end{table}

\begin{table}[tb]
  \caption{Maximum Hoop and Temperature for Candidate Cladding (LOCA)}
  \label{tbl:T_Hoop-Props2}
  \begin{center}
  \resizebox{\columnwidth}{!}{\begin{tabular}{l r r r r}
  \hline
   & Haynes 230  & Zircaloy-4  & SiC-SiC  & FeCrAl  \\
  \hline
  Fuel Max Temp (K) & 1183.39 & 1184.67 & 1173.13 & 1184.45  \\
  Clad Max Hoop (Pa) & 312901431.40 & 821237597.80 & 1285333500.00 & 383790824.10  \\
  Clad Max Temp (K)  & 1182.28 & 1183.63 & 1171.03 & 1183.39  \\
  \hline
  \end{tabular}}
  \end{center}
\end{table}

Heat generation within the fuel block (informed by Equation \ref{eq:APP}) and a convective heat flux boundary condition along the heat pipe walls (to simulate coolant) resulted in a temperature distribution for the model. This represented NOC. To simulate a LOCA in a heat pipe, the heat flux boundary would be removed for that respective pipe. Using MOOSE's Tensor Mechanics module, hoop stress in the ring clamp and heat pipe were calculated for both NOC and LOCA scenarios. The LOCA was simulated in only one heat pipe, located at the symmetrical boundary, for maximum angular visualization of its effects.

%\subsection{Results}
\label{subsec:MOOSER}

\begin{figure}[tb]
  \begin{center}
   \includegraphics[width=4.75in]{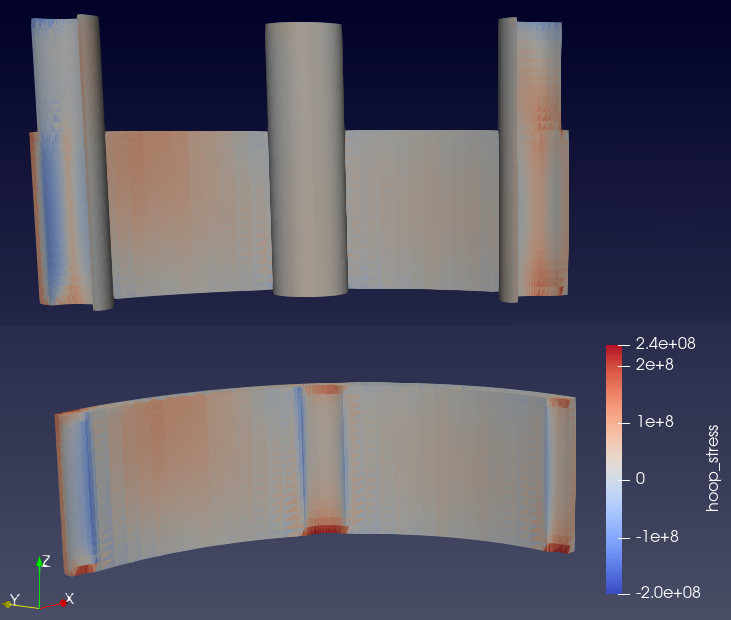}
  \end{center}
  \caption{Kilopower RCD section with Haynes 230 cladding hoop stress (Pa) during NOC, specifically the ring clamp and heat pipes.}
\label{fig:MOOSEFinalNOCExdStress1}
\end{figure}%

\begin{figure}[tb]
  \begin{center}
   \includegraphics[width=4.25in]{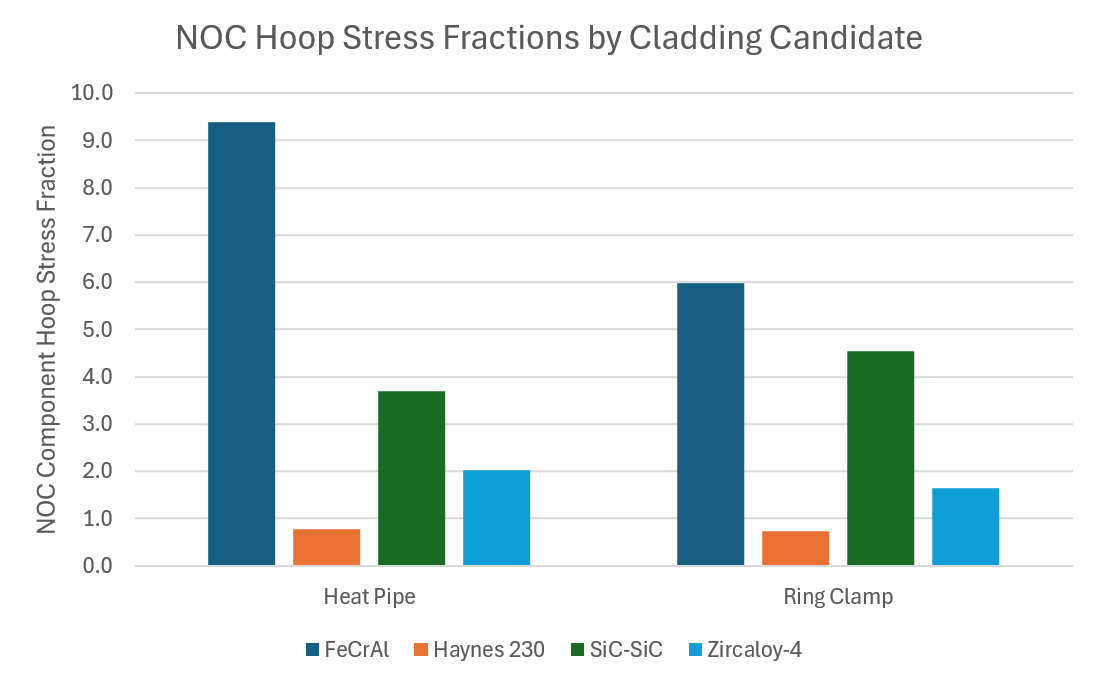}
  \end{center}
  \caption{Hoop stress fractions of the NOC runs, sorted by component and candidate cladding.}
\label{fig:NOCFracHoopInitial}
\end{figure}%

\begin{figure}[tb]
  \begin{center}
   \includegraphics[width=4.25in]{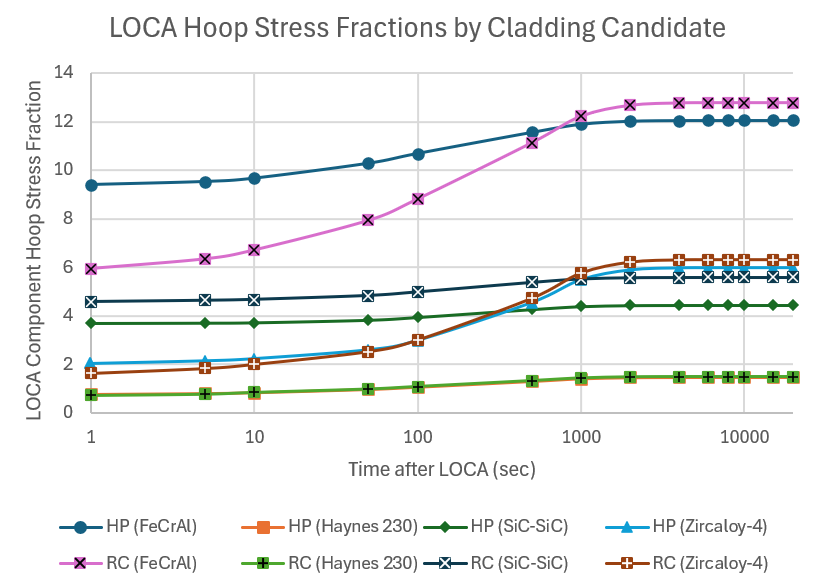}
  \end{center}
  \caption[Hoop stress fractions of the LOCA runs, sorted by component and candidate cladding.]{Hoop stress fractions of the LOCA runs, sorted by component and candidate cladding. HP is heat pipe and RC is ring clamp. HP (Haynes 230) and RC (Haynes 230) occupy nearly the same values.}
\label{fig:LOCAFracHoopInitial}
\end{figure}%

Across the heat pipe and ring clamp, the maximum stresses were measured using MOOSE’s Postprocessors module. Since maximum allowable stress is a scalar fraction of yield strength (YS) or fracture strength (FS) based on an arbitrary factor of safety, the maximum stresses found will be weighed against the YS or FS (whichever is lower) \cite{khattak.2019}. This is the fraction to dictate the candidates’ relative performance during NOC using Equation \ref{eq:Fraction}. It will be referred to as the ‘NOC [component] hoop stress fraction’ henceforth.

Maximal hoop stress varies wildly by candidate (Figure \ref{fig:MOOSEFinalNOCExdStress1}). Haynes 230 displays the lowest, almost matched by FeCrAl, while Zircaloy-4 nearly triples this and is only surpassed by SiC-SiC. Because the YS / FS limit depends closely upon temperatures for three of the four candidates - seen in Figure \ref{fig:Cladding-Limits} - the high temperatures of this reactor environment very harshly affected the NOC Heat Pipe and Ring Clamp stress fractions for FeCrAl and to a lesser extent Zircaloy-4. The sizeable difference between FeCrAl's component stress fractions is from a temperature-induced difference in limits, changing from 50 MPa for the HP to 30 MPa for the RC. SiC-SiC is a more unique case, for while it has a similar limit to Haynes 230, its own resistance to thermal creep is likely what produced extremely high stresses and rose its fraction so dramatically. The maximam hoop stresses for other candidates are also tabulated in Table \ref{tbl:T_Hoop-Props1} for NOC and Table \ref{tbl:T_Hoop-Props2} for LOCA. Maximum fuel temperatures and maximum cladding temperatures are also shown for all the candidates. Of all the candidates, Haynes 230 seems most well-equipped for this environment. Its limits are still relatively generous during these operational temperatures and its fractions are all lower than 1.0, meaning its maximum hoop stress and temperatures are less than their limits and theoretically will not have issues during NOC (Figure \ref{fig:NOCFracHoopInitial}). Like with NOC, the postprocessed output looked for maximum hoop stress across all elements of the ring clamp and heat pipes, with its own accompanying `LOCA [component] hoop stress fraction' (Equation \ref{eq:Fraction}).

As a LOCA is a time-dependent incident, where temperatures and stresses begin to increase after the accident occurs, these comparative measures tracked the maximums along each timestep. Factoring in materials stress limits with LOCA hoop stress fractions in Figure \ref{fig:LOCAFracHoopInitial} led to some radical effects on FeCrAl and especially Zircaloy-4 due to the slight change in temperature. While the absolute maximum Hoop Stress Fractions of SiC-SiC and Zircaloy-4 appear similar, the significant incremental increase in SiC-SiC’s stress fraction under LOCA conditions highlights its susceptibility to localized stress concentrations. Unlike metals, SiC-SiC’s ceramic nature and brittle fracture behavior make it more sensitive to such stress spikes, which could lead to sudden mechanical failure. This distinction is critical in evaluating accident-tolerant material performance beyond just maximum stress values. Yet even Haynes 230's LOCA stress fraction exceeds unity after a two-heat pipe LOCA, suggesting that even though it would be able to operate thermally, the ring clamps and heat pipes would be in danger of fracture. Haynes 230 performs the closest to within limits during a one-heat pipe LOCA, making it still the best choice for such a reactor in terms of accident prevention.

\subsection{Uncertainty Quantification and Sensitivity Analysis}
\label{sec:UQSA}

Although the results from MOOSE say much about the candidate claddings' performance, there is more analysis that can be done with the model. All M\&S have inherent uncertainty, originating from either unforeseen errors in the simulation (aleatoric) or inaccuracies in the model's setup, such as boundary conditions or other assumptions (epistemic). This paper will account for epistemic uncertainty in some candidate material properties using uncertainty quantification (UQ) and the effects of that variance on results will be measured with sensitivity analysis (SA). This has been performed via the coding language Python, notably the packages \emph{chaospy} and \emph{pandas} for their abilities to generate / sample distributions and read large volumes of data (specifically \emph{csv} files) respectively.

\begin{table}[tb]
  \caption{Expected RC Hoop Stress Fractions by Model}
  \label{tbl:Surrogate}
  \begin{center}
    \begin{tabular}{|l|r|r|r|r|}
      \hline
       & MOOSE & Surrogate & Surrogate Std Dev & Std Dev (\% of Surrogate) \\
      \hline
      NOC Haynes 230 & 0.73816 & 0.73892 & 0.03879 & 4.5369 \\
      NOC Zircaloy-4 & 1.6390  & 1.6458  & 0.09858 & 5.2499 \\
      NOC SiC-SiC    & 4.5494  & 4.5491  & 0.19852 & 4.3639 \\
      NOC FeCrAl     & 5.9798  & 5.9928  & 0.27189 & 5.9897 \\
      LOCA End Haynes 230 & 1.4900 & 1.4924 & 0.07668 & 4.4774 \\
      LOCA End Zircaloy-4 & 6.3172 & 6.3384 & 0.37044 & 5.1382 \\
      LOCA End SiC-SiC    & 5.5884 & 5.5879 & 0.24233 & 4.3368 \\
      LOCA End FeCrAl     & 12.793  & 12.803  & 0.57326 & 5.8443 \\
      \hline
    \end{tabular}
  \end{center}
\end{table}

%\subsection{Results}
\label{subsec:UQSAR}

\begin{figure}[tb]
  \begin{center}
   \includegraphics[width=4.5in]{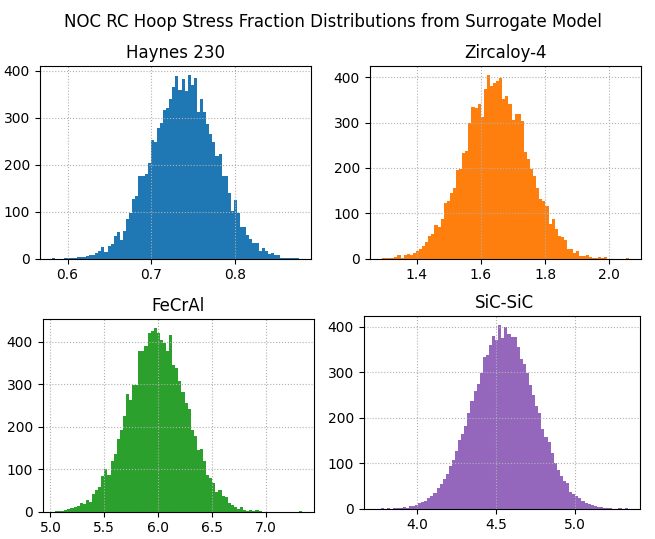}
  \end{center}
  \caption{Distributions of ring clamp hoop stress fractions (Histograms or frequencies in the Y axis and ring clamp hoop stress fraction values in the X axis) among the NOC surrogate model runs by candidate cladding.}
\label{fig:NOCSM}
\end{figure}%

\begin{figure}[tb]
  \begin{center}
   \includegraphics[width=4.5in]{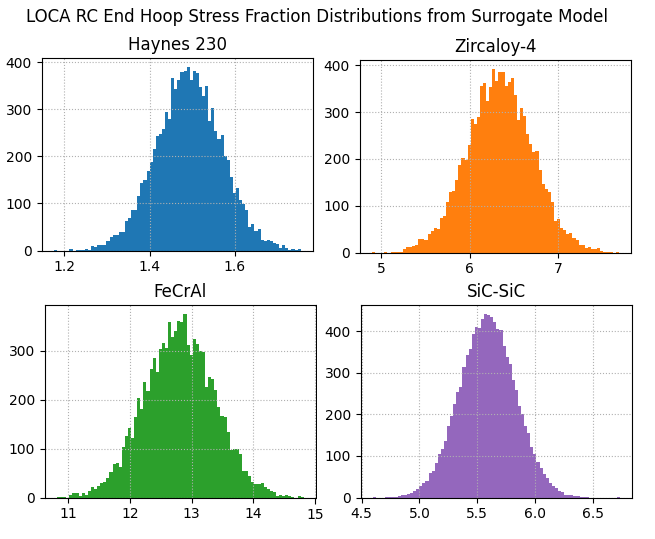}
  \end{center}
  \caption{Distributions of ring clamp end timestep hoop stress fractions (Histograms or frequencies in Y axis and ring clamp end timestep hoop stress fraction values in X axis) among the LOCA surrogate model runs by candidate cladding.}
\label{fig:LOCASM}
\end{figure}%

\begin{figure}[tb]
  \begin{center}
   \includegraphics[width=5.75in]{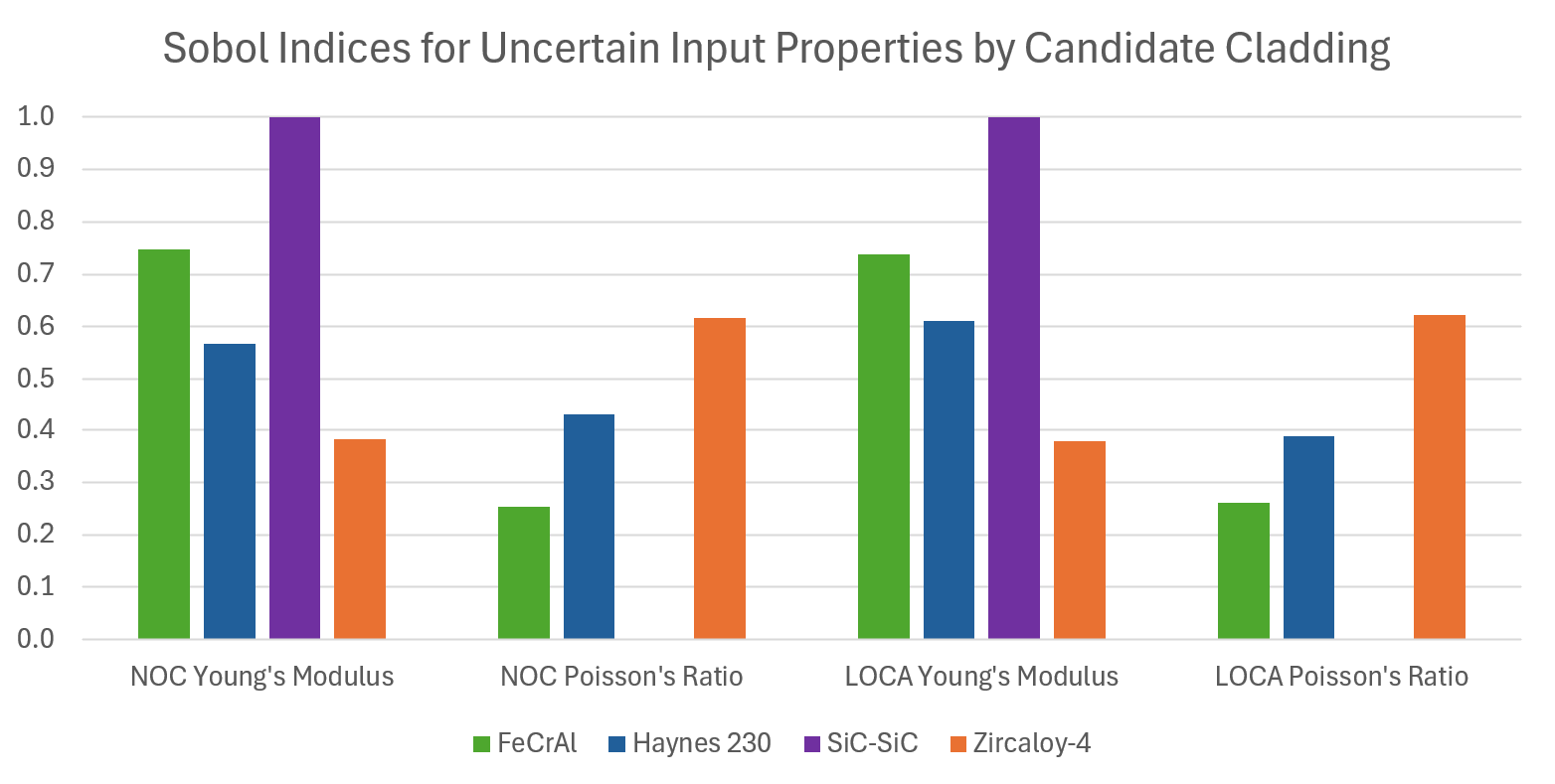}
  \end{center}
  \caption{Sobol indices of the uncertain input parameters for both NOC and LOCA, colour-coded by candidate cladding.}
\label{fig:Sobol}
\end{figure}%

Use of UQ and SA are recent trends in engineering \cite{kumar.2023}. They allow for even the most simple of M\&S efforts to still measure and account for variance in either the data being used or results generated \cite{kobayashi.2023}. This paper will use them to investigate how sensitive the MOOSE model's results are to uncertainty in some candidate claddings' material properties, by running it in large input decks. 

%\subsection{Methodology}
\label{subsec:UQSAM}

The method chosen is non-intrusive Polynomial Chaos Expansion (PCE) \cite{kobayashi2024ai,kumar.2021,kumar.2023}, which is sampling-based and data-driven. `Non-intrusive' means it does not require modification of the base M\&S to be performed. In this method, choice input variables of the model are assigned uncertainties, expressed as probability distributions. These inputs and all model outputs are represented by orthogonal polynomial coefficients of the PCE and can be used to compute statistical moments (such as mean and standard deviation) for the outputs using projection, regression, or collocation.

To start, the input variables are given PDFs and a surrogate model can be created: \begin{equation}\label{StochasticResponse}
    u(x;\xi) = \sum_{i=0}^{N} u_{i}(x)\psi_{i}(\xi)
\end{equation} 

A model with $n$ uncertain inputs has a stochastic output $u(x;\xi)$ that can be represented by Equation \ref{StochasticResponse}, where $u_i$ are the deterministic polynomial coefficients and $\psi_i$ are the multi-dimensional orthogonal polynomial functions of independent random variables $\xi = \{\xi_1, \xi_2, ..., \xi_n\}$. The PDF of $\xi$ is given by: \begin{equation}\label{InputPDF}
    \mathbf{W}(\xi) = \prod_{i=1}^{n} W_{i}(\xi_{i})
\end{equation} The distribution of $\xi$ is simply the joint probabilities of all individual $\xi_i$'s, represented by $W_i(\xi_i)$.

The sum of polynomial terms $N+1$ in Equation \ref{StochasticResponse} is in relation to the maximum order of the polynomial \emph{p} being expanded and the number of uncertain inputs \emph{n} as follows: \begin{equation}\label{SumPolyTerms}
    N+1 = \frac{(n+p)!}{n!p!}
\end{equation}

Because of the Kronecker delta $\sigma_{ij}$, polynomials $\psi_i$ are orthogonal as such: \begin{equation}\label{Ortho}
    \langle\psi_{i}\psi_{j}\rangle\ =\ \langle\psi_{i}^{2}\rangle\sigma_{ij}
\end{equation} $\langle\psi_i\psi_j\rangle$ is also defined as: \begin{equation}\label{Ortho2}
    \langle\psi_{i}\psi_{j}\rangle\ = \int_{\xi} \psi_{i}(\xi)\psi_{j}(\xi)\mathbf{W}_{\xi}d(\xi)
\end{equation}

Output $u$'s mean is found via: \begin{equation}\label{OutputMean}
    \langle u\rangle\ = \int_{\xi} \mathbf{W}_{\xi}d(\xi) = u_{0}
\end{equation} Due to orthogonality, PC coefficient $u_{0}$ is the mean of $u(x;\xi)$, the output response. The response's variance is computed by: \begin{equation}\label{OutputVariance}\begin{aligned}
    \sigma_{u}^{2} =\ \langle(u-\langle u\rangle)^{2}\rangle \\
    = \int_{\xi} \biggl\{ \sum_{i=1}^{N}u_{i}(x)\psi_{i}(\xi) \biggl\}^{2} \mathbf{W}_{\xi}d(\xi) \\
    = \sum_{i=1}^{N}u_{i}^{2}(x)\langle\psi_{i}^{2}\rangle
\end{aligned}\end{equation}

To perform UQ on output responses for inputs with any individually-given probability, the polynomials must be generalized such that: \begin{equation}\label{SystemOfEq}\begin{aligned}
    u(x;\xi^{1}) = u_{0} + u_{1}(x)\psi_{1}(\xi^{1}) + u_{2}(x)\psi_{2}(\xi^{1}) + ... + u_{N}(x)\psi_{N}(\xi^{1})\\
    u(x;\xi^{2}) = u_{0} + u_{1}(x)\psi_{1}(\xi^{2}) + u_{2}(x)\psi_{2}(\xi^{2}) + ... + u_{N}(x)\psi_{N}(\xi^{2})\\
    ...\\
    u(x;\xi^{M}) = u_{0} + u_{1}(x)\psi_{1}(\xi^{M}) + u_{2}(x)\psi_{2}(\xi^{M}) + ... + u_{N}(x)\psi_{N}(\xi^{M})
\end{aligned}\end{equation} can represent them as a system of equations using Equation \ref{StochasticResponse} as its base, $M$ being the total number of samples generated for $\xi$ ($\xi^{1}$, $\xi^{2}$, ... , $\xi^{M}$). $u(\xi^{i})$ are the corresponding model responses for each $\xi_{i}$. It can be rewritten in matrix form as: \begin{equation}\label{eq:MatrixOfEq}\begin{aligned}
\begin{pmatrix}
    \psi_{0}(\xi^{1}) & \psi_{1}(\xi^{1}) & \psi_{2}(\xi^{1}) & ... & \psi_{P}(\xi^{1})\\
    \psi_{0}(\xi^{2}) & \psi_{1}(\xi^{2}) & \psi_{2}(\xi^{2}) & ... & \psi_{P}(\xi^{2})\\
    ... & ... & ... & & ...\\
    \psi_{0}(\xi^{M}) & \psi_{1}(\xi^{M}) & \psi_{2}(\xi^{M}) & ... & \psi_{P}(\xi^{M})
\end{pmatrix}
\times
\begin{pmatrix}
    u_{0}\\
    u_{1}\\
    ...\\
    u_{P}
\end{pmatrix}
=
\begin{pmatrix}
    u(\xi^{1})\\
    u(\xi^{2})\\
    ...\\
    u(\xi^{M})
\end{pmatrix}
\end{aligned}\end{equation} or \begin{equation}\label{SimpleMatrix}\begin{aligned}
DU = B
\end{aligned}\end{equation} Should the sum $M$ of equations be higher than sum $P$ of unknowns, the solution of Equation \ref{eq:MatrixOfEq} can be found using least squares regression as such: \begin{equation}\label{eq:LSR}
    U = (D^{T}D)^{-1}D^{T}B
\end{equation}

This is the Wiener-Askey scheme using Legendre polynomials, which are most efficient for uniformly distributed random variables and as such are befitting of this study \cite{Wiener-Askey}. The system of equations (Equation \ref{SystemOfEq}) is solved using regression and its statistical moments (mean, standard deviation, etc.) are found as the final part of this step.

The PCE performed in this paper was to find the distributions of stress fractions during NOC and LOCA. Then, compare those distributions with the respective limits and with each other, finding which candidates' property uncertainties would have a higher likelihood of exceeding limits and thus would need to be prioritized for further experimental research. 50 runs of each configuration (NOC/LOCA and cladding candidate) were performed with 5\% uncertainty for both Young's Modulus and Poisson's Ratio of the cladding, with each run being a sample of these distributions. The resulting maximum hoop stresses were collected into their own distributions.

Next, this analysis sought to perform a robust SA on the PCE results using Sobol' sensitivity. Like non-intrusive PCE, Sobol' indices allow for a `black box' approach to the model input/output and don't require modification of the original M\&S. The indices are used to determine the influence of individual inputs on the variance of the output, with the sensitivity index $S_{j}$, written as: \begin{equation}\label{eq:LSR2}
    S_{j} = \frac{V_{j}}{V}
\end{equation} The sensitivity of the $j^{th}$ output parameter comes from total variance $V$ and $V_{j}$, the variance contribution from the $j^{th}$ input parameters.

Sobol' indices require user-defined distributions for input \emph{and} output to be sampled from. Since the M\&S software operates as a black box, its output distributions are undefined. This means code is needed to either define a distribution from the output data or create a surrogate model connecting the input and output samples (which represents the black box). Using \emph{chaospy}, the latter method was chosen. Several polynomial orders were tested, with first-order being the best representation when comparing the `expected' value from the surrogate model to the MOOSE model's results.

Following the PCE methodology previously outlined, surrogate models of each of the eight runs (four NOC, four LOCA) were created. These models were used for calculations of Sobol' indices (for the uncertain input parameters), as well as mean and standard deviation (of the output).

Taking 10,000 samples from each surrogate model gives Figures \ref{fig:NOCSM} and \ref{fig:LOCASM}. The `noise' in some of the sampling (for example, LOCA - FeCrAl compared to the smoother LOCA - SiC-SiC) is being accounted by the model, as repeated runs capture such behavior from the input distributions. This is an example of how \emph{chaospy} extrapolates behavior from the existing input-output data, making these surrogate models that much more robust. This likely originates in the Sobol' Indices for each cladding (Figure \ref{fig:Sobol}), where SiC-SiC hoop stress is revealed to be near-wholly dependent on Young's Modulus and not Poisson's Ratio. This one-variable dependency likely creates little cross-variable interaction in the polynomial and results in a more uniform distribution.

The tight relationship between the expected stress fractions of the MOOSE and surrogate model runs in Table \ref{tbl:Surrogate} represents the latter's accuracy of the original material. The standard deviation is a measure of the sensitivity of each configuration's maximum ring clamp hoop stress fraction. Its raw values are weighted by the expected values, so finding the standard deviation as a percentage was deemed more useful. These findings suggest that SiC-SiC and Haynes 230 exhibit more stable and predictable mechanical behavior, as their percent sensitivity is lower, indicating that small variations in material properties have a minimal impact on stress fractions. This reinforces their suitability for high-temperature applications, but further experimental testing is needed to validate these predictions. In contrast, Zircaloy-4 and FeCrAl show significantly higher sensitivity, implying that small changes in material properties result in large variations in stress response. This suggests a greater need for experimental validation of their high-temperature strength, oxidation resistance, and creep behavior to refine their suitability for accident-tolerant cladding applications. These results indicate that future work should prioritize reducing uncertainty in Zircaloy-4 and FeCrAl performance metrics, while for SiC-SiC and Haynes 230, the focus should be on long-term performance testing and validation in reactor environments.

Although the hoop stress fractions for FeCrAl are already \emph{much} higher than unity, meaning that efforts to find more definitive material properties will not move the needle much on their applicability to Kilopower-like reactors, Zircaloy-4 at NOC is close enough to warrant more material property investigations at higher temperatures. While lowering the Hoop Stress Fraction (HSF) below 1 under NOC through material modifications may improve Zircaloy-4’s performance, its significantly higher stress fractions under LOCA scenarios must also be taken into account. Even with improved material properties, Zircaloy-4 may still face structural limitations in accident scenarios, requiring further investigation into its viability as an accident-tolerant cladding material. Should the hoop stress fraction be ultimately lowered past 1, it may become a viable material for these applications.

\section*{Conclusion \& Future Work}\label{GPsec}
This study successfully replicated the multi-physics behavior of the NASA Kilopower reactor, enabling a comparative analysis of four candidate cladding materials under NOC and LOCA scenarios. The findings highlight the superior stress resistance of Haynes 230, the only material to remain within its yield strength during normal operation, making it the most promising candidate for high-temperature microreactor applications. While SiC-SiC exhibited excellent thermal resistance, its inherent limitations under mechanical stress, typical of ceramics, hinder its suitability for applications where structural integrity under high stress is critical. In contrast, Zircaloy-4 and FeCrAl demonstrated significantly higher hoop stress fractions, exceeding their allowable limits under LOCA conditions. These results discourage further investigation into Zircaloy-4 and FeCrAl for Kilopower-like reactors due to their inadequate high-temperature performance. However, Zircaloy-4's behavior in NOC conditions suggests that further characterization of its high-temperature properties could reveal its potential viability.

The study emphasizes the critical role of UQ in material performance evaluations. UQ is essential for understanding the variability in material properties and the impact of these variations on reactor performance, especially under extreme conditions. Through uncertainty quantification and sensitivity analysis, the study identifies the most influential material properties—such as Young’s Modulus and Poisson’s Ratio—whose uncertainty significantly affects the cladding's mechanical performance. This highlights the need for a robust database of high-temperature material properties, as the accuracy of modeling and simulation efforts depends on these values. Incorporating UQ ensures that the assessment of materials like Haynes 230, SiC-SiC, Zircaloy-4, and FeCrAl is based on deterministic properties and accounts for potential variations in real-world applications. As this study has shown, some materials, particularly Zircaloy-4 and FeCrAl, have high sensitivity to material property variations, suggesting that further experimental research is needed to refine their high-temperature and accident-tolerant behavior. The need for comprehensive UQ frameworks in nuclear reactor materials research is thus clear, as they provide insights into material vulnerabilities and inform decisions on which materials should undergo further testing and development.

The study also underscores the broader need for a more comprehensive database of high-temperature material properties to improve the accuracy of modeling and simulation for advanced reactors. Beyond thermomechanical performance, future research should consider material cost, manufacturability, and chemical compatibility to evaluate candidate cladding materials more holistically. The Kilopower reactor, as a validated and successfully tested system, serves as a valuable benchmark for future investigations. As NASA and private entities advance space exploration, the demand for Kilopower-like reactors will continue to grow, reinforcing the need for ongoing material development and refinement.

Future work will augment this study in the context of the intelligent digital twin, specifically leveraging the \textit{previous studies} of the research group on AI methods developed, such as Physics-regularized neural networks \cite{kobayashi2024physics}, multi-stage deep learning framework \cite{daniell2025digital}, deep neural operators \cite{kobayashi2024deep,kobayashi2024improved}, transfer learning \cite{kabir2024transfer}, explainability \cite{kobayashi2024explainable}. Moreover, AI interference will be designed to provide UQ on-the-fly built on previous studies \cite{kobayashi2024ai,kumar2019influence, kumar2022multi}. Integrating AI with multiphysics, real-time signal processing, and advanced/virtual sensors holds immense potential for advanced reactors. ML models trained on simulation and sensor data can predict performance, optimize designs, and act as virtual sensors \cite{hossain2025virtual}. This combination enables real-time system monitoring and continuous optimization \cite{sakib2011basic,hossain2024sensor, kabir2010theory,kabir2010watermarking}. Ultimately, this approach could improve heat pipe performance, enhance reactor reliability, and greater autonomy in advanced reactor operations \cite{kabir2010non,kabir2010loss}.

\section*{Data and code availability}
The data and code used and/or analyzed during this study are available from the corresponding author on reasonable request.

\section*{Competing Interests}
The authors declare no conflict of interest.

\section*{Author Contributions}
Alexander Foutch performed the analyses and wrote the manuscript. Kazuma Kobayashi and Dinesh Kumar helped in the UQ assessment. Ayodeji Alajo and Syed Bahauddin Alam helped in the analysis. Syed Bahauddin Alam supervised the project.

\section*{Declaration of Generative AI and AI-assisted technologies in the writing process}
During the preparation of this work the author(s) used ChatGPT in order to language editing and refinement. After using this tool/service, the author(s) reviewed and edited the content as needed and take(s) full responsibility for the content of the publication. [\href{https://www.elsevier.com/about/policies/publishing-ethics/the-use-of-ai-and-ai-assisted-writing-technologies-in-scientific-writing}{Elsevier Publishing Ethics}]

\bibliographystyle{unsrtnat}
\bibliography{references} 

\end{document}